\documentclass[twoside,twocolumn,9pt]{article}
\usepackage{extsizes}
\usepackage[super,sort&compress,comma]{natbib}
\usepackage[version=3]{mhchem}
\usepackage[left=1.5cm, right=1.5cm, top=1.785cm, bottom=2.0cm]{geometry}
\usepackage{balance}
\usepackage{times,mathptmx}
\usepackage{sectsty}
\usepackage{graphicx}
\usepackage{lastpage}
\usepackage[format=plain,justification=justified,singlelinecheck=false,font={stretch=1.125,small,sf},labelfont=bf,labelsep=space]{caption}
\usepackage{float}
\usepackage{fancyhdr}
\usepackage{fnpos}
\usepackage[english]{babel}
\addto{\captionsenglish}{%
  
}
\usepackage{array}
\usepackage{droidsans}
\usepackage{charter}
\usepackage[T1]{fontenc}
\usepackage[usenames,dvipsnames]{xcolor}
\usepackage{setspace}
\usepackage[compact]{titlesec}
\usepackage{hyperref}

\usepackage{latexsym}
\usepackage{siunitx}
\usepackage{longtable}
\usepackage{booktabs}
\usepackage{multirow}
\makeatletter
\let\oldlt\longtable
\let\endoldlt\endlongtable
\def\longtable{\@ifnextchar[\longtable@i \longtable@ii}
\def\longtable@i[#1]{\begin{figure}[t]
\onecolumn
\begin{minipage}{0.5\textwidth}
\oldlt[#1]
}
\def\longtable@ii{\begin{figure}[t]
\onecolumn
\begin{minipage}{0.5\textwidth}
\oldlt
}
\def\endlongtable{\endoldlt
\end{minipage}
\twocolumn
\end{figure}}
\makeatother

\definecolor{cream}{RGB}{222,217,201}

\begin{document}

\pagestyle{fancy}
\thispagestyle{plain}
\fancypagestyle{plain}{

\fancyhead[C]{\includegraphics[width=18.5cm]{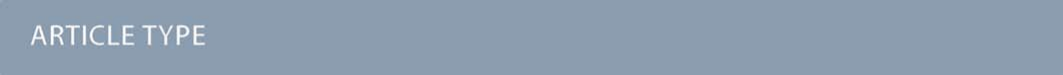}}
\fancyhead[L]{\hspace{0cm}\vspace{1.5cm}\includegraphics[height=30pt]{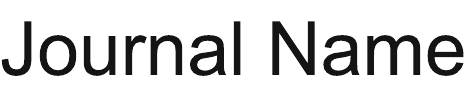}}
\fancyhead[R]{\hspace{0cm}\vspace{1.7cm}\includegraphics[height=55pt]{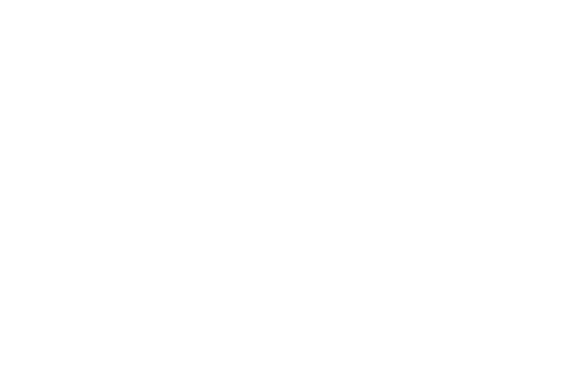}}
\renewcommand{\headrulewidth}{0pt}
}

\makeFNbottom
\makeatletter
\renewcommand\LARGE{\@setfontsize\LARGE{15pt}{17}}
\renewcommand\Large{\@setfontsize\Large{12pt}{14}}
\renewcommand\large{\@setfontsize\large{10pt}{12}}
\renewcommand\footnotesize{\@setfontsize\footnotesize{7pt}{10}}
\makeatother

\renewcommand{\thefootnote}{\fnsymbol{footnote}}
\renewcommand\footnoterule{\vspace*{1pt}%
\color{cream}\hrule width 3.5in height 0.4pt \color{black}\vspace*{5pt}}
\setcounter{secnumdepth}{5}

\makeatletter
\renewcommand\@biblabel[1]{#1}
\renewcommand\@makefntext[1]%
{\noindent\makebox[0pt][r]{\@thefnmark\,}#1}
\makeatother
\renewcommand{\figurename}{\small{Fig.}~}
\sectionfont{\sffamily\Large}
\subsectionfont{\normalsize}
\subsubsectionfont{\bf}
\setstretch{1.125} 
\setlength{\skip\footins}{0.8cm}
\setlength{\footnotesep}{0.25cm}
\setlength{\jot}{10pt}
\titlespacing*{\section}{0pt}{4pt}{4pt}
\titlespacing*{\subsection}{0pt}{15pt}{1pt}

\fancyfoot{}
\fancyfoot[LO,RE]{\vspace{-7.1pt}\includegraphics[height=9pt]{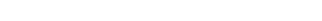}}
\fancyfoot[CO]{\vspace{-7.1pt}\hspace{13.2cm}\includegraphics{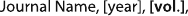}}
\fancyfoot[CE]{\vspace{-7.2pt}\hspace{-14.2cm}\includegraphics{head_foot/RF}}
\fancyfoot[RO]{\footnotesize{\sffamily{1--\pageref{LastPage} ~\textbar  \hspace{2pt}\thepage}}}
\fancyfoot[LE]{\footnotesize{\sffamily{\thepage~\textbar\hspace{3.45cm} 1--\pageref{LastPage}}}}
\fancyhead{}
\renewcommand{\headrulewidth}{0pt}
\renewcommand{\footrulewidth}{0pt}
\setlength{\arrayrulewidth}{1pt}
\setlength{\columnsep}{6.5mm}
\setlength\bibsep{1pt}

\makeatletter
\newlength{\figrulesep}
\setlength{\figrulesep}{0.5\textfloatsep}

\newcommand{\topfigrule}{\vspace*{-1pt}%
\noindent{\color{cream}\rule[-\figrulesep]{\columnwidth}{1.5pt}} }

\newcommand{\botfigrule}{\vspace*{-2pt}%
\noindent{\color{cream}\rule[\figrulesep]{\columnwidth}{1.5pt}} }

\newcommand{\dblfigrule}{\vspace*{-1pt}%
\noindent{\color{cream}\rule[-\figrulesep]{\textwidth}{1.5pt}} }

\makeatother

\twocolumn[
  \begin{@twocolumnfalse}
\vspace{3cm}
\sffamily
\begin{tabular}{m{4.5cm} p{13.5cm} }

\includegraphics{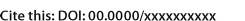} & \noindent\LARGE{\textbf{Non-collinear
magnetism in the post-perovskite thiocyanate frameworks
CsM(NCS)\(_3\)}} \\
\vspace{0.3cm} & \vspace{0.3cm} \\

 & \noindent\large{Madeleine Geers,\textit{$^{a, b}$} Jie Yie
Lee,\textit{$^{a, b}$} Sanliang Ling,\textit{$^{c}$} Oscar
Fabelo,\textit{$^{b}$} Laura Cañadillas-Delgado,\textit{$^{b}$} Matthew
J.
Cliffe,$^{\ast}$\textit{$^{a}$} $^{\dag}$} \\
\includegraphics{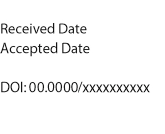} & \noindent\normalsize{\ce{AMX3}
compounds are structurally diverse, a notable example being the
post-perovskite structure which adopts a two-dimensional framework with
corner- and edge-sharing octahedra. Few molecular post-perovskites are
known and of these, none have reported magnetic structures. Here we
report the synthesis, structure and magnetic properties of molecular
post-perovskites: \ce{CsNi(NCS)3}, a thiocyanate framework, and two new
isostructural analogues \ce{CsCo(NCS)3} and \ce{CsMn(NCS)3}.
Magnetisation measurements show that all three compounds undergo
magnetic order. \ce{CsNi(NCS)3} (Curie temperature,
\(T_\mathrm{C} = 8.5(1)\;\)K) and \ce{CsCo(NCS)3}
(\(T_\mathrm{C} = 6.7(1)\;\)K) order as weak ferromagnets. On the other
hand, \ce{CsMn(NCS)3} orders as an antiferromagnet (Néel temperature,
\(T_\mathrm{N}=16.8(8)\;\)K). Neutron diffraction data of
\ce{CsNi(NCS)3} and \ce{CsMn(NCS)3}, show that both are non-collinear
magnets. These results suggest molecular frameworks are fruitful ground
for realising the spin textures required for the next generation of
information
technology.} \\

\end{tabular}

 \end{@twocolumnfalse} \vspace{0.6cm}

  ]

\renewcommand*\rmdefault{bch}\normalfont\upshape
\rmfamily
\section*{}
\vspace{-1cm}

           \footnotetext{\textit{$^{a}$~School of Chemistry, University
Park, Nottingham, NG7 2RD, United
Kingdom~Email: matthew.cliffe@nottingham.ac.uk}}
               \footnotetext{\textit{$^{b}$~Institut Laue Langevin, 71
avenue des Martyrs CS 20156, 38042 Grenoble Cedex 9, France}}
               \footnotetext{\textit{$^{c}$~Advanced Materials Research
Group, Faculty of Engineering, University of Nottingham, University
Park, Nottingham NG7 2RD, United Kingdom}}
    
\footnotetext{\dag~Electronic Supplementary Information (ESI) available: Additional
experimental details for synthesis, single crystal X-ray and neutron
diffraction measurements and analysis, powder neutron diffraction
measurements and analysis, magnetic measurements, density functional
theory calculations; CIFs and mCIFs.. See DOI: 00.0000/00000000.}



\hypertarget{introduction}{%
\section{Introduction}\label{introduction}}

A unifying goal in solid-state science is control over the physical
properties of materials, and the tunability of perovskites is perhaps
the most striking example. Traditionally these compounds are
three-dimensional frameworks with a general chemical formula \ce{AMX3}
built from \ce{[MX6]} corner-sharing octahedra. Through substitution of
the A, M and X ions, materials with remarkable
magnetic,\citep{vishnoi_structural_2020} electronic
conductivity,\citep{jaffe_high_2020}
photovoltaic\citep{kojima_organometal_2009} and non-linear optical
properties,\citep{xu_halide_2020} have been created.

\begin{figure}[!ht]
  \includegraphics[width=8.4cm]{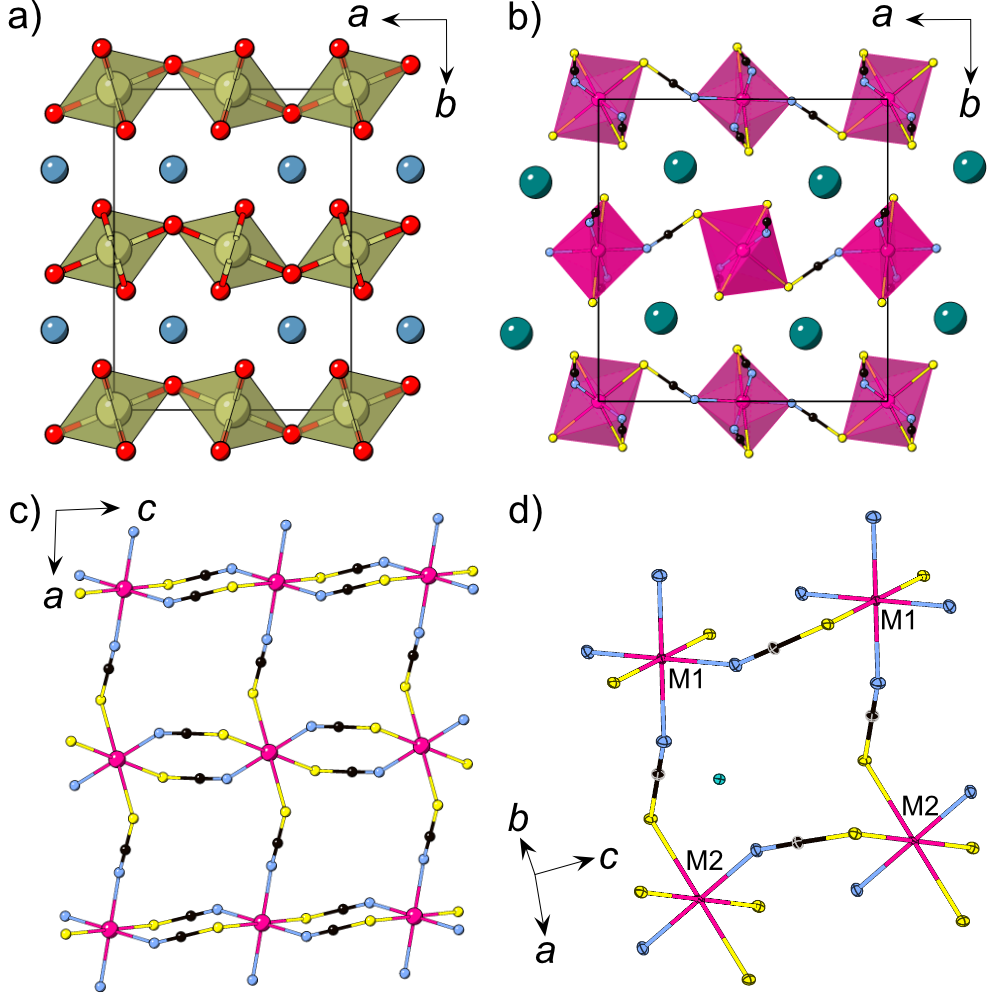}
  \caption{a) The structure of \ce{CaIrO3} with iridium polyhedra connected $via$ corner-sharing ($a$ axis) and edge-sharing ($c$ axis) \ce{[IrO6]} octaheadra, Ca = blue, Ir = brown, O = red. b) Crystal structure of \ce{CsMn(NCS)3} at 120 K obtained from single crystal X-ray diffraction. c) A single \ce{[Mn(NCS)3]-} layer. d) Close-up view of of the structure highlighting the two chemically independent metal ions (M1, M2). Cs = green, Mn = pink, N = blue, C = black, S = yellow.}
  \label{fig:structure4}
\end{figure}

The perovskite structure is, however, only one of a wide-range of
\ce{AMX3} structure-types, with perhaps the most closely related being
the post-perovskite structure. In contrast to perovskites,
post-perovskites contain \ce{[MX6]} octahedra that both edge- and
corner-share, resulting in two-dimensional anionic \ce{[MX3]} layers,
rather than a three dimensional framework. These layers are stacked with
interstitial A cations positioned between them. Post-perovskites are
amongst the most abundant terrestrial minerals, as the \ce{MgSiO3}
perovskite that makes up much of the Earth's lower mantle undergoes a
critical high-pressure and temperature phase transition
(\(P_C \approx 125 \;\)GPa, \(T_C \approx 1250\;\)K) to the
post-perovskite structure-type near the mantle-core
boundary.\citep{oganov_theoretical_2004, murakami_post-perovskite_2004}
Due to the difficulty of reaching these extreme pressures,
post-perovskites which form at more accessible pressures and can be
recovered on quenching have proved useful analogues. These include the
second- and third-row transition metal oxides \ce{AMO3}, A = Na, Ca, and
M = Pt, Rh, Ir
(\(P_\mathrm{syn} \approx 5\;\)GPa);\citep{ohgushi_capto3_2008, yamaura_synthesis_2009, rodi_ternare_1965, bremholm_nairo3pentavalent_2011}
and first-row fluorides \ce{NaMF3}, M = Mg, Ni, Co, Fe,
Zn.\citep{martin_rietveld_2006, shirako_high-pressure_2012, dobson_towards_2011, Lindsay-scott_time_2014, bernal_perovskite_2014}
A handful of post-perovskite compounds can even be obtained at ambient
conditions: notably \ce{CaIrO3},\citep{ohgushi_resonant_2013} the
post-actinide chalcogenides \ce{AMnSe3} (A = Th,
U)\citep{ijjaali_syntheses_2004} and
\ce{UFeS3},\citep{noel_structure_1976} and
\ce{TlPbI3}.\citep{lin_inorganic_2021} As a result of this synthetic
challenge, systematic tuning of the properties of post-perovskites is
much less well explored than for perovskites.

In particular, there are limits on our current understanding of the
magnetic properties of post-perovskites, in part because neutron
diffraction studies require large sample sizes. This is despite the fact
that post-perovskites, unlike perovskites, tend to have non-collinear
magnetic
structures.\citep{bogdanov_post-perovskite_2012, ohgushi_resonant_2013, shirako_high-pressure_2012, bernal_perovskite_2014}
As a result, both the exploration of fundamental properties of
post-perovskites and the potential utility of their non-trivial spin
textures for spintronic
devices\citep{jia_persistent_2020, tokura_magnetic_2021} or quantum
memory storage\citep{li_high-performance_2021} remains limited.

In contrast to the relative scarcity of atomic post-perovskites,
molecular post-perovskites, where X is a molecular anion, are a growing
class of
materials.\citep{raebiger_1-d_2001, van_der_werff_cation_2001, Fleck2004, maczka_temperature-dependent_2018, wang_temperature-induced_2019}
Unlike their atomic analogues, the majority of molecular
post-perovskites are stable and synthesisable at ambient pressure.
Incorporating molecular components in these frameworks can also allow
for novel physical properties, arising from the additional degrees of
freedom, including non-linear optics,\citep{Triki2020} electric
polarisation \citep{Bostrom2018, Lee2021} and complex spin
textures.\citep{lou_tunable_2020}

Metal dicyanamides, \ce{AM(dca)3} dca = \ce{N(CN)2-}, are the best
established family of molecular
post-perovskites.\citep{biswas_retention_2006, wang_two-dimensional_2015, wang_unique_2019}
The metal ions are separated by five-atom-bridges and tend to be
magnetically isolated, with no conclusive evidence of long-range
magnetic
ordering.\citep{van_der_werff_cation_2001, van_der_werff_structure_2001, raebiger_1-d_2001}
To explore collective magnetic behaviour in molecular post-perovskite
analogues we therefore focussed on ligands capable of propagating
stronger superexchange interactions.

Thiocyanate (\ce{NCS-}) is a promising molecular ligand for creating
magnetic coordination frameworks with long-range magnetic ordering, even
over extended
distances.\citep{Bassey2020, Cliffe2018, cliffe_magnetic_2022, baran_neutron_2019, defotis_antiferromagnetism_1993, Shurdha2012}
The thiocyanate also has asymmetric reactivity, unlike both atomic
ligands and most common molecular ligands, \emph{e.g.} formate or
\ce{dca-}.\citep{boca_polymethylammonium_2004} Applying the Hard-Soft
Acid-Base principle,\citep{Pearson1963} the nitrogen terminus is less
polarisable and so coordinates preferentially to harder first-row
transition metals, whereas sulfur has more diffuse orbitals, and
preferentially coordinates to softer main group second- and third-row
transition metals.\citep{Cliffe2018} Homoleptic framework structures are
thus comparatively rare, beyond the binary
\ce{M(NCS)2}.\citep{Cliffe2018, Bassey2020} Nevertheless, there are two
reported \ce{AM(NCS)3} homoleptic frameworks with the post-perovskite
structure: \ce{RbCd(NCS)3}\citep{Thiele1980} and
\ce{CsNi(NCS)3}.\citep{Fleck2004}

In this work, we study \ce{CsM(NCS)3}, M = Ni, Mn and Co.~We describe
the synthesis of \ce{CsMn(NCS)3} and \ce{CsCo(NCS)3} and determine their
structures, by single crystal X-ray diffraction, to be post-perovskites
isomorphic with \ce{CsNi(NCS)3}. We used bulk magnetometry to show that
all three order magnetically. \ce{CsNi(NCS)3}
(\(T_\mathrm{C} = 8.5(1)\;\)K) and \ce{CsCo(NCS)3}
(\(T_\mathrm{C} = 6.7(1)\;\)K) order as weak ferromagnets with
significant coercive fields, \ce{CsNi(NCS)3}
\(H_\mathrm{C} = 0.331(2)\;\)T and \ce{CsCo(NCS)3}
\(H_\mathrm{C} = 0.052(2)\;\)T, whereas \ce{CsMn(NCS)3} orders as an
antiferromagnet \(T_\mathrm{N}=16.8(8)\;\)K. We then report neutron
diffraction measurements, with single crystal and powder samples. These
found \ce{CsNi(NCS)3} to be a canted ferromagnet,
\(\mathbf{k} = (0,0,0)\), and \ce{CsMn(NCS)3} to be a non-collinear
antiferromagnet which orders with
\(\mathbf{k} = (0, \frac{1}{2}, \frac{1}{2})\) and four sublattices
which all order antiferomagnetically. Additionally, density functional
theory (DFT) calculations were undertaken, which confirm the intralayer
interactions are at least an order of magnitude stronger than the
interlayer interactions in these compounds.

Our results suggest that magnetic molecular frameworks, and thiocyanates
in particular, are a promising ground for exploring the magnetism of
otherwise challenging to realise structure types. The non-collinear
structures we have uncovered in these thiocyanate frameworks \emph{via}
neutron diffraction suggest that further studies of this family may
uncover new routes to complex spin textures.

\hypertarget{results}{%
\section{Results}\label{results}}

\hypertarget{synthesis-and-structure}{%
\subsection{Synthesis and Structure}\label{synthesis-and-structure}}

We synthesised \ce{CsNi(NCS)3}, \ce{CsMn(NCS)3} and \ce{CsCo(NCS)3} by
salt metathesis of the metal sulfate and caesium sulfate with barium
thiocyanate in aqueous solution. \ce{CsNi(NCS)3} is stable at ambient
conditions, however \ce{CsMn(NCS)3} and \ce{CsCo(NCS)3} are sensitive to
humidity. Single crystals were obtained through slow evaporation and the
structures were determined using single crystal X-ray diffraction to
have the monoclinic space group \(P2_1/n\), and to crystallise in the
post-perovskite structure. We found that, unlike the Ni and Mn
analogues, the crystallisation of \ce{CsCo(NCS)3} from solution
typically occurs in a two step process. Large, deep blue crystals of a
second phase, believed to be \ce{Cs2Co(NCS)4}, are typically obtained,
which recrystallise into small, deep purple single crystals of
\ce{CsCo(NCS)3} if left undisturbed for several weeks in ambient
conditions.

All three compounds are isomorphic and consist of anionic
{[}\ce{M(NCS)3}{]} layers in the \emph{ac} plane in which the transition
metal M\(^{2+}\) ions are connected through \(\mu_{1,3}\)NCS ligands
(Fig.~\ref{fig:structure4}\(\mathrm{c}\)). In between the layers, which
are stacked along the \emph{b} axis, lie the charge balancing caesium
counterions. The M\(^{2+}\) ions are octahedrally coordinated and there
are two crystallographically and chemically distinct metal sites. One
transition metal ion (M1, Wyckoff site 2\emph{c}) is coordinated by four
nitrogen atoms and two sulfur atoms, whilst the second transition metal
ion (M2, Wyckoff site 2\emph{b}) is bonded to four sulfur and two
nitrogen atoms. The metal octahedra corner-share along the \emph{a} axis
and alternate between M1 and M2. Along the \emph{c} axis, the metal
octahedra edge-share, and all the metal sites within an edge-sharing
chain are the same.

\begin{figure*}[!ht]
  \hypertarget{fig:magnetometry2}{%
  \centering
  \includegraphics[width=17.4cm]{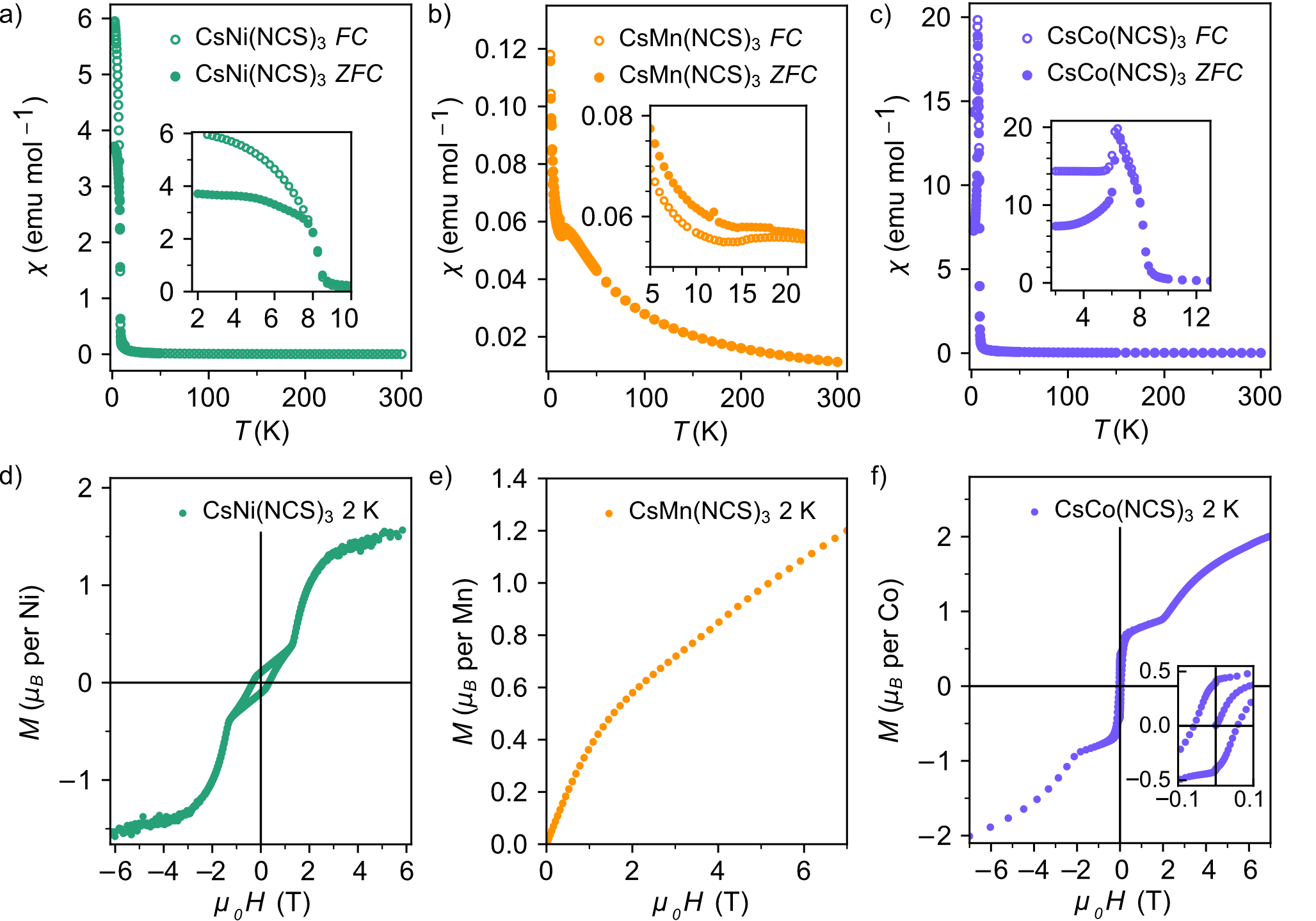}
  \caption{Magnetic susceptibility data with an applied field of $0.01$ T for a) \ce{CsNi(NCS)3}; b) \ce{CsMn(NCS)3}; c) \ce{CsCo(NCS)3}; and isothermal magnetisation measurements at 2 K for d) \ce{CsNi(NCS)3} (field $-6$ to $+6$ T); e) \ce{CsMn(NCS)3} (field $0$ to $+7$ T); and f) \ce{CsCo(NCS)3} (field $-7$ to $+7$ T).}\label{fig:magnetometry2}
  } 
\end{figure*}

Moving along the row from manganese to nickel, the average \ce{M-N} and
\ce{M-S} bond lengths, \(d_\mathrm{M-N}\) and \(d_\mathrm{M-S}\),
decrease: \(d_\mathrm{Mn-N} (13 \; \mathrm{K}) = 2.1607(14)\;\)Å,
\(d_\mathrm{Co-N} (120 \; \mathrm{K}) = 2.072(3)\;\)Å,
\(d_\mathrm{Ni-N} (15 \; \mathrm{K}) = 2.035(5)\;\)Å and
\(d_\mathrm{Mn-S} = 2.674(3)\;\)Å, \(d_\mathrm{Co-S} = 2.5652(10)\;\)Å,
\(d_\mathrm{Ni-S} = 2.350(20)\;\)Å. The \ce{M-S} bond lengths shorten to
a greater extent than the \ce{M-N} bond lengths, likely as a result of
the higher polarisability of sulfur. This trend is comparable in other
metal thiocyanate frameworks.\citep{Bassey2020, Palion-Gazda2015} There
are two intralayer M\(\cdots\)M distances, between the edge-sharing
octahedra (\(d_\mathrm{M1-M1}\) = \(d_\mathrm{M2-M2} = c\)) and the
corner-sharing octahedra (\(d_\mathrm{M1-M2}= a\)). Again, moving from
\ce{Mn$^{2+}$} to \ce{Ni$^{2+}$}, the distance decrease, with
\(d_\mathrm{Mn1-Mn1} = 5.6540(4)\;\)Å compared to
\(d_\mathrm{Co1-Co1} = 5.57860(11)\;\)Å and
\(d_\mathrm{Ni1-Ni1} = 5.5409(6)\;\)Å; and
\(d_\mathrm{Mn1-Mn2}= 6.37315(5)\;\)Å,
\(d_\mathrm{Co1-Co2} = 6.30515(5)\;\)Å and
\(d_\mathrm{Ni1-Ni2} = 6.2631(6) \;\)Å. The interlayer spacing also
decreases, with the shortest M\(\cdots\)M distances,
\(d_\mathrm{M,layer}\), decreasing from
\(d_\mathrm{Mn,layer} = 7.20052(15)\) Å, through
\(d_\mathrm{Co,layer} = 7.18340(10)\) Å to
\(d_\mathrm{Ni,layer} = 7.1050(8)\) Å. For the single crystal neutron
diffraction measurements of \ce{CsNi(NCS)3} a significantly smaller
crystal was used in comparison to the \ce{CsMn(NCS)3}, which influences
the quality of the reflections, resulting in the larger errors.

\begin{table}
\caption{DFT predicted relative energies between the experimental post-perovskite phase (\ce{CsM(NCS)3} = pPv) and the two known \ce{AM(NCS)3} perovskite structure types: \ce{CsCd(NCS)3} = \ce{Cs[Cd]}; \ce{(NH4)2NiCd(NCS)6} = \ce{NH4[NiCd]}. The perovskite structures were generated by swapping the A site cation for \ce{Cs+} and the B site cation(s) by the appropriate transition metal.}
\label{table:DFT}
\begin{tabular}{lrrr}
\ce{M^{2+}}     &  $E$(pPv)$^\ast$   &  $E$(\ce{Cs[Cd]})$^\ast$ &  $E$(\ce{NH4[NiCd]})$^\ast$ \\
\hline \\
Ni              &        0.0              & $+90.8$    &      $+177.3$    \\
Mn              &        0.0              &  $+84.8$   &      $+137.0$    \\
Co              &        0.0              & $+286.7$   &      $+114.5$    
\end{tabular}

$ \ast$ (meV per formula unit)

\end{table}

In addition to this post-perovskite structure, there are two different
perovskite structure-types with composition \ce{AM(NCS)3},
\ce{CsCd(NCS)3}\citep{Thiele1980} and \ce{(NH4)2NiCd(NCS)6} (ESI Fig.
7).\citep{Xie2016} To understand the relative stability of the
post-perovskite structure compared to the perovskite-type structures,
density functional theory (DFT) calculations were performed.
Hypothetical perovskites were generated through atom-swaps, and the
geometry optimised relaxation, relative to the experimentally observed
post-perovskite structure, was calculated (Table \ref{table:DFT}). For
simplicity, we focused on the spin-ferromagnetic solution of the three
different structure types for this comparison. We found that the
post-perovskite structure was more stable than the perovskite structures
by around 10 kJ mol\(^{-1}\) (0.1 eV per formula unit, approximately
\(4\;kT\) at room temperature). This energy is consistent with the
observed exclusive formation of the post-perovskite structure, but
suggests that more unusual experimental conditions may allow access to
the perovskite phases.

\hypertarget{magnetism}{%
\subsection{Magnetism}\label{magnetism}}

\begin{figure*}[!ht]
\centering
  \includegraphics[width=17.4cm]{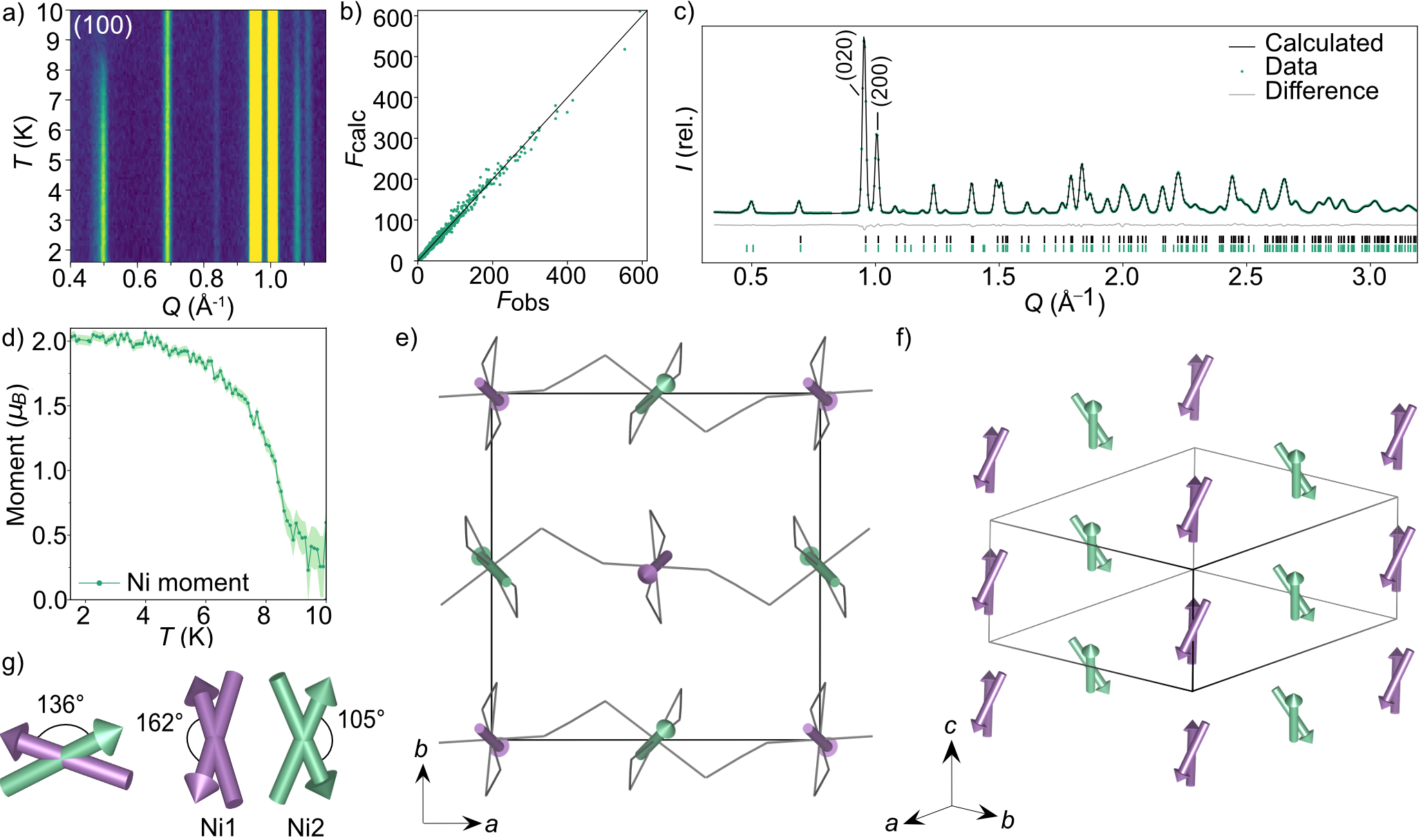}
  \caption{Neutron diffraction data and the magnetic structure for \ce{CsNi(NCS)3}. a) Thermodiffractogram measured between 1.5 and 10 K on the D1b diffractometer (ILL). The most intense magnetic Bragg reflection is indexed as the (100) planes. The $F_{\mathrm{obs}}$ against $F_{\mathrm{calc}}$ plot obtained from D19 data fit (nuclear and magnetic) collected at 1.8 K (b) and Rietveld fit obtained from D1b data fit at 1.5 K (c) for the multi pattern refinement of the $P2_1/c$ magnetic model describing the magnetic structure of \ce{CsNi(NCS)3}. d) The magnetic moment of the Ni$^{2+}$ ions as a function of temperature obtained by Rietveld refinements of the data collected at each temperature point between 1.5 and 10 K. e) The magnetic structure of \ce{CsNi(NCS)3} viewed down the $c$ axis. The two unique magnetic vectors are represented with purple arrows (Ni1) and green arrows (Ni2). The \ce{Cs+} cations have been omitted for clarity and the thiocyanate ligands are represented as a wire frame. f) The magnetic structure viewed down the [111] direction. g) Angles describing the non-collinearity of the moments.}
  \label{fig:Ni_neutron}
\end{figure*}

\begin{figure*}[!ht]
\centering
  \includegraphics[width=17.4cm]{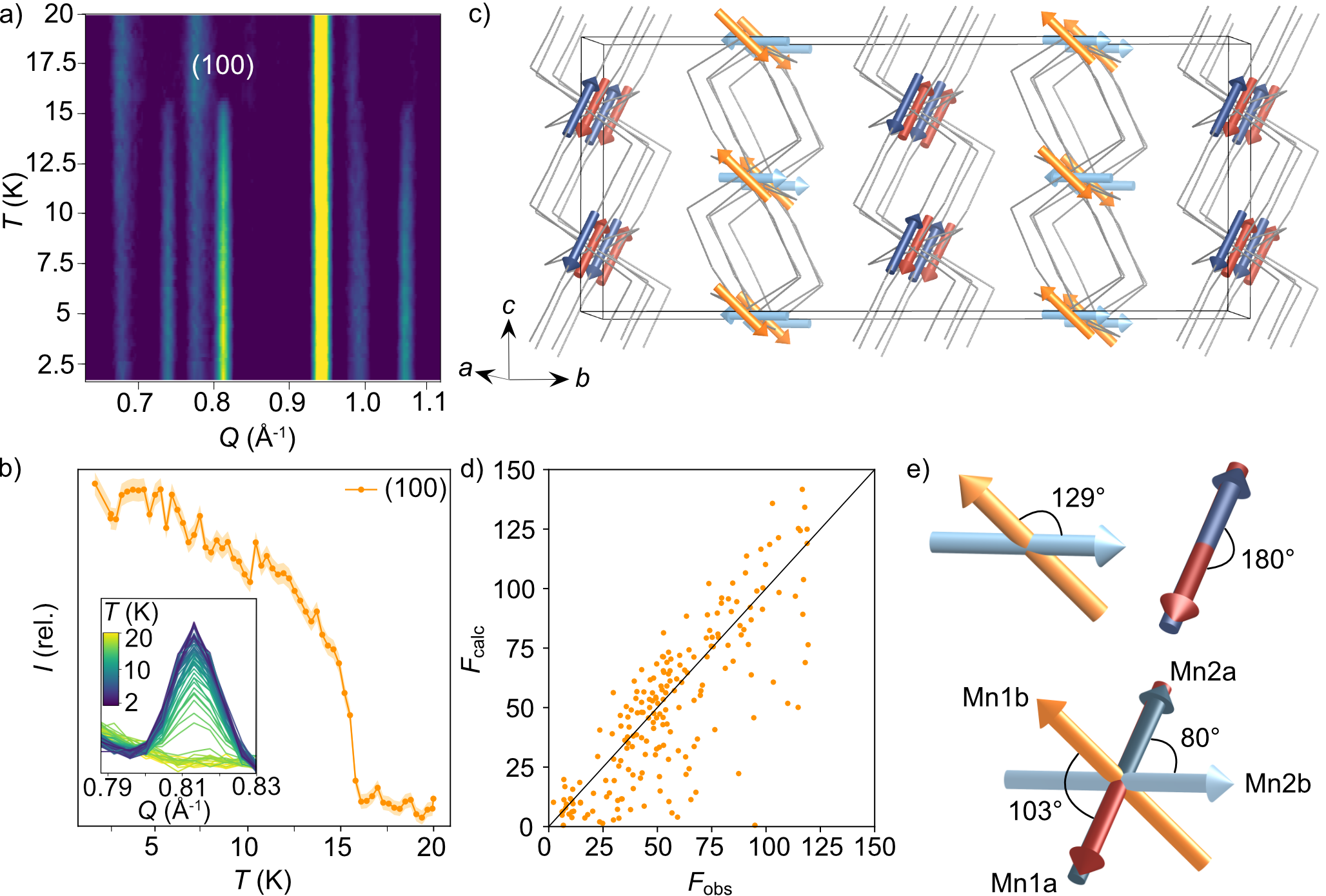}
  \caption{Neutron diffraction data used to determine the magnetic structure for \ce{CsMn(NCS)3}. a) Thermodiffractogram measured between 1.5 and 20 K on the D1b diffractometer (ILL). The most intense magnetic Bragg reflection is indexed as the (100) planes. b) The integrated intensity of the (100) magnetic reflection at 0.81 $\mathrm{\mathring{A}^{-1}}$ as a function of temperature. c) The magnetic structure of \ce{CsMn(NCS)3}. The magnetic vectors are depicted with red and orange arrows for Mn1 (red = Mn1a and orange = Mn1b) and blue arrows for Mn2 (dark = Mn2a and light = Mn2b). The \ce{Cs+} cations have been omitted for clarity and the thiocyanate ligands are represented as a wire frame. d) $F_\mathrm{obs}$ against $F_\mathrm{calc}$ for the refinement of the $P_S\overline{1}$ magnetic model using the magnetic reflections observed at 2 K with the D19 diffractometer (ILL). e) The angles between the magnetic vectors.}
  \label{fig:Mn_neutron}
\end{figure*}

Having synthesised an isostructural series of thiocyanate
post-perovskites containing paramagnetic ions, we next sought to
understand their magnetic behaviour. We measured the bulk magnetic
susceptibility of each of these compounds, in an applied field of
\(0.01\) T, to determine the magnetic ordering temperature and average
interaction strength. We used isothermal magnetisation measurements at 2
K over the range \(-7\) to \(+7\) T (\(-6\) to \(+6\) T for M = Ni) to
assess the degree of magnetic hysteresis.

The zero-field cooled (\emph{ZFC}) and field-cooled (\emph{FC})
susceptibility of \ce{CsNi(NCS)3} diverge below the ordering temperature
\(T_\mathrm{C} = 8.5(1)\;\)K
(Fig.~\ref{fig:magnetometry2}\(\mathrm{a}\)). The Curie-Weiss fit of the
inverse susceptibility above \(180\) K gives a value for the Curie-Weiss
temperature, \(\theta_\mathrm{CW}\), of \(-8.6(8)\) K. However, this
value is particularly sensitive to the fitting temperature range:
\(\theta_\mathrm{CW} = +1.5(4)\) K when \(100< T < 300\;\)K, whereas
\(\theta_\mathrm{CW}\) = \(-12.7(8)\) K for a fit \(200 < T < 300\;\) K.
This variation is likely due to the presence of significant single-ion
anisotropy, typical of
\ce{Ni$^{2+}$}.\citep{liu_magnetic_2018, curley_magnetic_2021} The Curie
constant, \(C\), is \(0.85(2)\) emu K mol\(^{-1}\), which is lower than
the spin only value, \(C_\mathrm{spin \; only} = 1\) emu K mol\(^{-1}\).
The lower than expected Curie constant (ESI Fig. 1) is likely due to a
sample mass error. The isothermal magnetisation of \ce{CsNi(NCS)3} at 2
K shows hysteresis with a coercive field of
\(H_\mathrm{C} = 0.331(2)\;\)T and a remnant magnetisation
\(M_\mathrm{rem} = 0.106(1) \; \mu_B\) per Ni, implying a canting angle
of 6.1° if there are only two distinct spin orientations. Beyond
\(1.19(1)\) T, the hysteresis loop closes and there is a magnetic phase
transition to a second magnetic phase, reaching a magnetisation of
\(1.54(1) \; \mu_B\) per Ni at 6 T.

The magnetic susceptibility of \ce{CsMn(NCS)3} has a cusp at \(16.8(8)\)
K indicating the onset of antiferromagnetic order
(Fig.~\ref{fig:magnetometry2}\(\mathrm{b}\)). An increase in
susceptibility at low temperature is likely due to a small fraction of a
hydrated impurity. Fitting the inverse susceptibility between
\(100<T<300\) K to the Curie-Weiss law gives
\(\theta_\mathrm{CW} =-33.6(2)\) K, and \(C = 3.75(2)\) emu K
mol\(^{-1}\), which is lower than the high-spin spin only expected value
of \(C_\mathrm{spin \; only} = 4.375\) emu K mol\(^{-1}\) for
\ce{Mn$^{2+}$}. We did not observe any evidence of hysteresis in the
isothermal magnetisation data, consistent with antiferromagnetic order.
The magnetisation reaches \(1.20(6) \; \mu_B\) per Mn at the largest
field measured (\(7.00(1)\) T), well below the spin only saturation
magnetisation, \(M_\mathrm{sat.} = 5\; \mu_B\) per Mn, indicative of the
presence of significant antiferromagnetic interactions. The ratio of the
Curie-Weiss temperature to the ordering temperature is
\(f = |\theta_\mathrm{CW} / T_\mathrm{N}| = 2.1\), suggestive of slight
frustration or low-dimensionality.

Curie-Weiss fitting of the magnetic susceptibility of \ce{CsCo(NCS)3}
between \(100<T<300\;\)K suggests predominately antiferromagnetic
interactions, \(\theta_\mathrm{CW}\) = \(-19.7(2)\) K. The large Curie
constant, \(C = 4.8(2)\) emu K mol\(^{-1}\), compared to the high-spin
spin only value \(C_\mathrm{spin \; only} = 1.875\) emu K mol\(^{-1}\),
indicates that the unquenched orbital moment remains significant in
these compounds and the determined \(\theta_\mathrm{CW}\) therefore
likely also includes the effects of the first order spin-orbit coupling.
d\(\chi\)/d\emph{T} shows two sharp minima, suggesting that there are
potentially two ordering temperatures for the compound at \(6.7(1)\) and
\(8.4(1)\) K. The isothermal magnetisation data measured at 2 K show a
hysteresis with \(H_\mathrm{C} = 0.052(2)\;\)T and a remnant
magnetisation \(M_\mathrm{rem} = 0.400(1) \; \mu_B\) per Co, suggesting
a canting angle of \(15.3\)°. The hysteresis disappears at \(1.86(1)\)
T, when \(M=0.88(4)\;\mu_B\) per Co.~Above this applied magnetic field,
the magnetisation steadily increases, reaching \(2.00(7)\; \mu_B\) per
Co at \(7.00(1)\) T, although the moment remains unsaturated, due to a
combination of single-ion anisotropy and antiferromagnetic interactions.

Our bulk magnetic properties measurements suggested that both
\ce{CsNi(NCS)3} and \ce{CsCo(NCS)3} are weak ferromagnets (canted
antiferromagnets), or more generally have uncompensated magnetic
moments, with appreciable hystereses and field-induced magnetic phase
transitions. Comparatively, the absence of hysteresis and a negative
\(\theta_\mathrm{CW}\) for \ce{CsMn(NCS)3} suggests it is an
antiferromagnet.

\hypertarget{neutron-diffraction}{%
\subsection{Neutron Diffraction}\label{neutron-diffraction}}

To explore the magnetic properties of these post-perovskites further, we
therefore carried out neutron diffraction experiments, both single
crystal and powder, to determine their ground state magnetic structures.
Scale-up of our initial synthetic routes produced high quality powder
and single crystal samples of \ce{CsNi(NCS)3} and \ce{CsMn(NCS)3}
suitable for neutron diffraction, however the two-stage synthesis of
\ce{CsCo(NCS)3} meant we were unable to obtain either large (mm\(^3\))
single crystals or gram-scale pure phase microcrystalline powders,
precluding neutron studies for this compound.

We carried out single crystal neutron diffraction measurements of
\ce{CsNi(NCS)3} (\(1.8\times0.9\times0.3\) mm\(^3\)) using the D19
diffractometer at the Institut Laue Langevin (ILL). A low temperature
data collection at 2 K, below the ordering temperature of 8.5 K, allowed
us to determine the propagation vector to be \(\mathbf{k} = (0, 0, 0)\)
by indexing of the magnetic Bragg reflections. We combined these single
crystal neutron diffraction (SCND) data with additional powder neutron
diffraction (PND) data collected on a polycrystalline sample (1.1 g)
using the powder neutron diffractometer D1b (ILL). The magnetic space
groups with maximal symmetry which permit magnetic moments to exist on
the two \ce{Ni^{2+}} ions were determined using the Bilbao
Crystallographic
Server\citep{aroyo_crystallography_2011, gallego_magnetic_2012, perez-mato_symmetry-based_2015}
to be \(P2_1'/c'\) and \(P2_1/c\) (BNS
notation).\citep{gallego_magnetic_2012} Both of these models have the
same unit cell as the nuclear structure. We then refined models in each
space group against the combined PND and SCND data with a multipattern
refinement and found that only \(P2_1/c\) was able to model the
additional intensity arising from the magnetic reflections. This refined
structure shows weak ferromagnetic order with two unique magnetic sites,
corresponding to the two crystallographic \ce{Ni$^{2+}$} ions (Ni1 and
Ni2) in the nuclear structure. The two moments have a magnitude of
\(2.01(3)\;\mu_B\) and were constrained to refine together with a
negative correlation. The moments are directed predominately along the
\emph{c} axis with the canting only present along the \emph{b} axis. The
Ni1 moments (purple arrows Fig.~\ref{fig:Ni_neutron}) are canted at an
angle of \(162\)° along the \(-b\) direction, whilst the Ni2 moments
(green arrows Fig.~\ref{fig:Ni_neutron}) are canted at an angle of
\(105\)° in the \(+b\) direction. The asymmetric canting of the two
sublattices results in a net magnetisation of \(0.116 \;\mu_B\) per Ni
along the \(b\) axis (\(+b\) direction). This is equivalent to a single
Ni\(^{2+}\) canting at an angle of \(6.7\)°, which is in agreement with
the magnetometry data. The magnetic moment of Ni1 and Ni2 present a
relative tilt of \(136\)° between them.

We found that when subtracting the 2 K PND data from the 10 K data,
there are some nuclear peaks which do not directly overlap at the two
temperatures. This is particularly evident in the (020) and (200)
reflections, which are the most intense in the diffraction patterns
(Fig.~\ref{fig:Ni_neutron}\(\mathrm{c}\)). On heating from 2 to 10 K,
the \emph{a} axis increases by \(0.028\) \%, whereas the \emph{b} axis
decreases in length by \(0.019\) \%. In contrast, the \emph{c} axis
remains constant within this temperature range (ESI Fig. 6).

We were also able to measure a large single crystal of \ce{CsMn(NCS)3}
(\(6.3\times2.5\times1.1\) mm\(^3\)) using D19 (ILL). We found in our
diffraction data collected at 2 K, below \(T_\mathrm{N}\) = 16.8 K, a
number of additional Bragg reflections (2938 unique reflections) not
present in our data collected at 20 K, which could not be indexed to the
nuclear structure. We were able to index these additional magnetic Bragg
reflections with \(\mathbf{k} = (0, \frac{1}{2}, \frac{1}{2})\). Using
the Bilbao Crystallographic Server we determined the possible magnetic
space groups for this propagation vector to be \(P_S\overline{1}\) and
\(P_S1\). We solved the magnetic structures in both magnetic space
groups, with the \(P_S\overline{1}\) better fitting the experimental
data. The model comprises of four unique magnetic sites: Mn1a and Mn1b,
which arise from nuclear site Mn1 but are in alternate layers (red and
orange arrows Fig.~\ref{fig:Mn_neutron}\(\mathrm{c}\)), and Mn2a and
Mn2b from nuclear site Mn2 (dark and light blue arrows
Fig.~\ref{fig:Mn_neutron}\(\mathrm{c}\)). The magnetic unit cell is
related to the nuclear cell as follows
\(a_\mathrm{mag} = a_\mathrm{nuc}\),
\(b_\mathrm{mag} = 2b_\mathrm{nuc}\) and
\(c_\mathrm{mag} = 2c_\mathrm{nuc}\). The magnitude of the moments for
all Mn sites is \(4.63(9) \; \mu_B\) and were constrained to refine with
a single moment value. We found that allowing the moments to refine
freely did not significantly improve our fit (constrained refinement,
\(\chi ^2 = 68.5\) compared to free refinement, \(\chi ^2 = 56.7\)), and
led to unphysically small moment sizes and unstable moment angles.
Constraining the moment angles of Mn1b and Mn2b to be collinear (while
allowing Mn1a and Mn2a to be non-collinear) gave \(\chi ^2 = 120.3\),
whilst constraining all four moments to be collinear gave
\(\chi ^2 = 807.6\). In the determined model, each of the four magnetic
sublattices, derived from a unique Mn\(^{2+}\) site, order
antiferromagnetically as expected from the bulk antiferromagnetic order
observed in the magnetisation data. Mn1a and Mn2a moments, are aligned
antiparallel within the anionic layer, while Mn1b and Mn2b are at an
angle of \(130\)° relative to each other
(Fig.~\ref{fig:Mn_neutron}\(\mathrm{e}\)). Powder neutron diffraction
data were collected on the D1b diffractometer, which clearly shows the
additional magnetic Bragg reflections. However, the limited data quality
prevented quantitative refinement of these data.

\hypertarget{density-functional-theory-calculations}{%
\subsection{Density Functional Theory
Calculations}\label{density-functional-theory-calculations}}

\begin{figure}
\hypertarget{fig:J_definitions}{%
\centering
\includegraphics{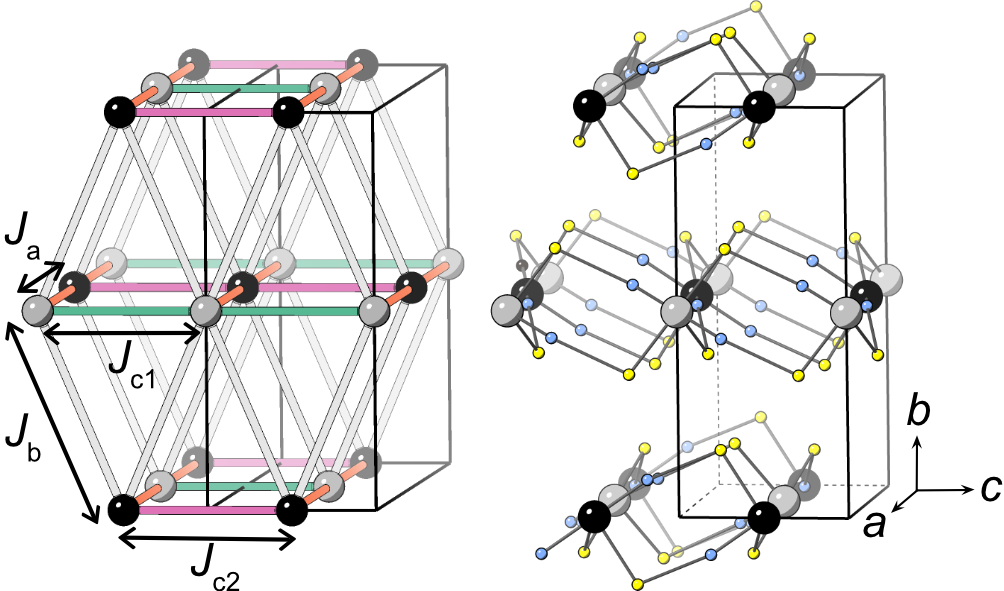}
\caption{The four nearest-neighbour magnetic interactions in
\ce{CsM(NCS)3}. M1 and M2 are grey and black respectively and the
thiocyanate ligands are represented with sulfur (yellow) and nitrogen
(blue) atoms only. Caesium cations and carbon atoms have been removed
for clarity.}\label{fig:J_definitions}
}
\end{figure}

Our neutron analysis provided the ground states, but the energetics
which yield these states remained opaque. We therefore turned to DFT
calculations. In order to correct for the typical delocalisation errors
encountered in DFT, a Hubbard \(U\) was
incorporated.\citep{Cliffe2018, wu_nincs2_2021} There are four
nearest-neighbour interactions (Fig.~\ref{fig:J_definitions}): \(J_a\),
through the M1\ce{-NCS-}M2 corner-sharing bridge; \(J_{c1}\), through
the M1\ce{-NCS-}M1 edge-sharing chain; \(J_{c2}\), through the
M2\ce{-NCS-}M2 edge-sharing chain; and \(J_b\), between the layers. The
magnetic lattice therefore consists of rectangular lattices in which
there are two different kinds of chain along one direction, which are
then coupled together by \(J_b\). Due to the offset of the layers, this
means that if \(J_a\), \(J_{c1}\) and \(J_{c2}\) are antiferromagnetic,
we would expect this lattice to be frustrated. To calculate each of
these interactions we therefore constructed eight \(2\times1\times1\)
supercells with distinct ordering patterns (ESI Table 1), and fitted
their DFT+\(U\) energies to the following Hamiltonian:
\begin{equation} \label{eq:hamiltonian}
E = \sum_{ij} J_{ij} S_{i} \cdot S_j + E_0
\end{equation}

where \(J_{ij}\) denotes the superexchange interaction for the atom pair
\(ij\) (i.e.~each interaction is counted once), \(|S|=1\), and \(E_0\)
is the energy of a hypothetical non-magnetic state. We found that for
\ce{CsNi(NCS)3} and \ce{CsMn(NCS)3} we could obtain self-consistent
results, but we were unable to achieve acceptable self-consistency for
\ce{CsCo(NCS)3}. The residual unquenched orbital moment for the
\(^4T_{1}\) ground state is likely responsible for this and suggests
that higher level calculations are likely to be required to
appropriately capture the magnetic behaviour of this compound.

Our calculations indicate that all four interactions are
antiferromagnetic for \ce{CsMn(NCS)3}, with the interactions through the
edge-sharing chains (\(J_{c1}\) and \(J_{c2}\)) stronger than the
corner-sharing bridge (\(J_a\)) between them (Table \ref{tab:DFT_mag}).
The magnitude of the \(J_b\) interaction is much smaller than the error
and is thus not meaningful. We found that \ce{CsNi(NCS)3} has
ferromagnetic interactions within the edge-sharing chains, with the
\(J_{c2}\) chain notably stronger, and the interactions between chains
are antiferromagnetic. As with \ce{CsMn(NCS)3}, the \(J_b\) interaction
is small and zero within error. From these interactions we were able to
calculate Curie-Weiss temperatures: for Mn
\(J_\mathrm{CW, calc.} = -22.3(5)\;\)K and for Ni
\(J_\mathrm{CW, calc.} = -0.05(65)\;\)K which are broadly comparable
with those found experimentally. The ground states predicted by DFT are
largely consistent with those determined experimentally, allowing for
the fact that these non-relativistic calculations cannot predict
spin-canting.

\begin{table}[]
\caption{DFT-derived superexchange energy, $J_x$, as defined in Eqn.~\ref{eq:hamiltonian} for \ce{CsM(NCS)3}, M = Ni, Mn.}
\label{tab:DFT_mag}
\begin{tabular}{lrrrr}
M$^{2+}$ & $J_a^\ast$ & $J_b^\ast$      & $J_{c1}^\ast$    & $J_{c2}^\ast$         \\
\hline \\
Ni       & 0.19(4)                   & $-$0.02(10)                   & $-$0.13(7)                     & $-$0.25(7)                     \\
Mn       & 0.78(3)                   & 0.03(6)                      & 2.40(6)                        & 1.79(6)                           
\end{tabular}

$^\ast$ (meV)

\end{table}

\hypertarget{discussion}{%
\section{Discussion}\label{discussion}}

The \ce{CsM(NCS)3} compounds all crystallise with the post-perovskite
structure. However, unlike most atomic analogues, these thiocyanate
compounds adopt the structure at ambient pressure. A common
characteristic of atomic post-perovskites is the presence of large
octahedral tilt angles in the corresponding perovskite phase, which
leaves them more susceptible to undergo the post-perovskite phase
transition.\citep{martin_rietveld_2006} Thiocyanate perovskites are
already very tilted, due to the shape of the frontier bonding orbitals,
which could explain the ease for formation of this structure type for
\ce{CsM(NCS)3}.\citep{Cliffe2018} Defining the plane as the \(ac\) axes,
the tilting occuring in these thiocyanate compounds along the edge- and
corner-sharing directions can be compared. The corner-sharing octahedra,
along the \emph{a} axis, have smaller deviations away from the \(ac\)
plane, with angles of \(\angle ac-\mathrm{M}-\mathrm{S} = 42(1)\)° and
\(\angle ac-\mathrm{M}-\mathrm{N} = 9(1)\)° (ESI Fig. 9). Along the
\emph{c} axis, the edge-sharing octahedra adopt greater tilting angles
as an inherent consequence of having two thiocyanate ligands bridging
each pair of metal centres in this direction. The deviation away from
the \(ac\) plane is \(\angle ac-\mathrm{M}-\mathrm{S} = 58(1)\)° and
\(\angle ac-\mathrm{M}-\mathrm{N} = 36(1)\)°. The incorporation of a
molecular ligand permits access to this greater degree of tilting
without the need for external pressure. This is evident from our DFT
calcualtions which show that for these \ce{CsM(NCS)3} compounds the
post-perovskite structure-type is lower energy than other reported
thiocyanate perovskite-type structures (Table \ref{table:DFT}).

Our magnetometry and neutron diffraction measurements show that the
compounds magnetically order between 6 and 16 K, significantly lower
than the closest chemical analogues, the binary thiocyanates
\ce{M(NCS)2} M = Mn, Fe, Co, Ni, Cu,\citep{Cliffe2018, Bassey2020} which
order at \(T_\mathrm{N}=29\;\)K for \ce{Mn(NCS)2},
\(T_\mathrm{N}=20\;\)K for \ce{Co(NCS)2}\citep{Shurdha2012, Bassey2020}
and \(T_\mathrm{N}=54\;\)K for
\ce{Ni(NCS)2}.\citep{Bassey2020, defotis_antiferromagnetism_1993} The
atomic post-perovskites have a range in ordering temperatures, with the
fluorides ordering at similar temperatures to \ce{CsM(NCS)3}, for
example post-perovskite \ce{NaNiF3} orders at \(T_\mathrm{N}=22\;\)K
(compared to \(T_\mathrm{C}=156\;\)K for the perovskite
phase).\citep{shirako_high-pressure_2012} The reported ordering
temperatures of the oxide post-perovskites are an order of magnitude
larger, for example \ce{CaIrO3} has
\(T_\mathrm{N}=115\;\)K,\citep{ohgushi_metal-insulator_2006, yamaura_synthesis_2009, shirako_magnetic_2010}
likely as the oxides lie closer to the metal-insulator boundary.

One key difference between \ce{CsM(NCS)3} and \ce{M(NCS)2} is that the
post-perovskites only have three-atom connections between transition
metals (\(\mu_{13}\)NCS coordination mode), whereas \ce{M(NCS)2} have
both one-atom and three-atom connections (\(\mu_{133}\)NCS). The
additional \ce{M-S-M} superexchange pathway in the binary thiocyanates
likely strengthens the magnetic interactions in the binary thiocyanates,
although DFT calculations of \ce{Cu(NCS)2} suggest that interactions
through the \ce{M-S-C-N-M} can be as strong or stronger than through
\ce{M-S-M} bridges.\citep{Cliffe2018} Our DFT calculations further
support this, showing appreciable superexchange through the end-to-end
bridging thiocyanates.

In contrast to these compounds, \ce{dca^-} based post-perovskites
containing magnetic ions do not appear to
order.\citep{van_der_werff_cation_2001, van_der_werff_structure_2001, raebiger_1-d_2001}
The transition metals in these compounds are separated by six bonds and
\(d\)(Mn-NCNCN-Mn) = 8.9825(4) Å
(\ce{[Ph4P]Mn(dca)3}),\citep{raebiger_1-d_2001} compared to four bonds
and \(d\)(Mn-NCS-Mn) = 6.37315(5) Å (\ce{CsMn(NCS)3}). This likely
reduces the superexchange further. However, \ce{Cr[Bi(SCN)6]}, with even
longer superexchange pathways does order at
\(T_\mathrm{N} = 4.0\;\)K,\citep{cliffe_magnetic_2022} indicating that
orbital overlap and orbital energy matching are also playing a key role
in this.

\ce{CsM(NCS)3}, M = Ni, Mn, Co, all adopt non-collinear magnetic
structures. Non-collinearity also appears to be typical in the atomic
perovskites. The only previous experimentally reported magnetic
structure of a post-perovskite is of \ce{CaIrO3}, which used magnetic
resonant X-ray scattering to reveal a canted stripe antiferromagnetic
ground state.\citep{ohgushi_resonant_2013} Octahedral tilting is
predicted to be a key parameter in determining the degree of
non-collinearity,\citep{weingart_noncollinear_2012} so it is expected
that the post-perovskite structures are sensitive to this factor as
well. Analysis of our isothermal magnetisation data for \ce{CsNi(NCS)3}
(Fig.~\ref{fig:magnetometry2}\(\mathrm{d}\)), assuming that there is
only a single magnetic site (i.e.~only two distinct spin orientations),
gives a canting angle of \(6\)°. However our magnetic structure has two
magnetic sites (i.e.~four spin orientations), and so there are in fact
three `canting angles', all of which are larger than \(6\)° (\(9\)°,
\(22\)°, and \(38\)°). The symmetry constraints of the \(P2_1/c\)
magnetic space group means that for each pair of canted moments (i.e.~a
single magnetic site), the components of the magnetic moments along the
\emph{a} and \emph{c} axes will be of equal magnitude and so the
uncompensated magnetisation lies only along the \emph{b} axis. In
\ce{CsNi(NCS)3}, the uncompensated moments from each magnetic site have
opposite signs: \(+0.80 \; \mu_B\) per Ni2 and \(-0.57 \; \mu_B\) per
Ni1, with a net moment of \(+0.114 \; \mu_B\). Using bulk measurements
for materials with complex magnetic structures can therefore lead to
underestimates of the degree of non-collinearity.

The magnetic structure of \ce{CsMn(NCS)3}, unlike the nickel and cobalt
analogues, orders as an antiferromagnet. The neutron data reveal a
\textbf{k} = \((0,\frac{1}{2}, \frac{1}{2})\) propagation vector, which
leads to four unique sublattices. As a result of the anticentring
translation in the \(P_S\overline{1}\) magnetic space group, each
sublattice, and therefore the overall structure, is antiferromagnetic.
There are two distinct kinds of layer within the magnetic structure, `a'
and `b'. Layer a, containing Mn1a and Mn2a, is antiferromagnetically
correlated; but Mn1b and Mn2b are non-collinear both with respect to
each other and also to Mn1a and Mn2a. The complexity of this structure
is perhaps surprising, considering the relative simplicity of the
nuclear structure and the lack of spin-orbit coupling expected for high
spin Mn\(^{2+}\). The layers stack so that each consecutive layer is
offset by \(c_\mathrm{nuc}/2\), resulting in a triangular relationship
between the interlayer moments (Fig.~\ref{fig:J_definitions}). This
layered stacking pattern may generate frustration, which could explain
the observed non-collinear structure.

\hypertarget{conclusion}{%
\section{Conclusion}\label{conclusion}}

In this paper, we have reported the synthesis, structure, magnetometry,
single crystal and powder neutron diffraction data for three isomorphic
post-perovskite thiocyanate frameworks, \ce{CsM(NCS)3} M = Ni, Mn,
Co.~Our magnetic susceptibility measurements show that all the materials
magnetically order, \ce{CsNi(NCS)3} \(T_\mathrm{C} = 8.5(1)\;\)K,
\ce{CsMn(NCS)3} \(T_\mathrm{N} = 16.8(8)\;\)K and \ce{CsCo(NCS)3}
\(T_\mathrm{C} = 6.7(1)\;\)K. Our neutron diffraction experiments on
\ce{CsNi(NCS)3} and \ce{CsMn(NCS)3} revealed both compounds have complex
non-collinear ordering.

\ce{CsNi(NCS)3} orders as a weak ferromagnet with two magnetically
distinct nickel moments. \ce{CsMn(NCS)3}, on the other hand, orders as
an antiferromagnet with a magnetic unit cell which is doubled along the
nuclear \emph{b} and \emph{c} axes, and has four unique sublattices.

Our neutron diffraction studies have shown that despite the relative
simplicity of the chemical structures, these thiocyanate
post-perovskites are a rich source of unusual magnetic orderings which
cannot be recognised through magnetometry data alone. Introducing
molecular ligands into framework structures may therefore provide a host
of unexpected and complex spin textures, motivating both future
synthetic and neutron diffraction investigations.

\section*{Conflicts of interest}
No conflicts of interest to declare.

\section*{Acknowledgements}
M.G. acknowledges the ILL Graduate School for provision of a
studentship. M.J.C. acknowledges the School of Chemistry, University of
Nottingham for support from Hobday bequest. We acknowledge the ILL for
beamtime under proposal numbers
5-41-1060,\citep{canadillas-delgado_ill_nodate}
5-12-344,\citep{canadillas-delgado_ill_nodate-1}
5-31-2726,\citep{canadillas-delgado_ill_nodate-2}
5-31-2816.\citep{canadillas-delgado_ill_nodate-3} Raw data sets from ILL
experiments can be accessed via links provided in references. Magnetic
measurements were carried out in part using the Advanced Materials
Characterisation Suite, funded by EPSRC Strategic Equipment Grant
EP/M000524/1. S.L. acknowledges the use of the Sulis supercomputer
through the HPC Midlands+ Consortium, and the ARCHER2 supercomputer
through membership of the U.K.'s HPC Materials Chemistry Consortium,
which are funded by EPSRC Grant numbers EP/T022108/1 and EP/R029431/1,
respectively.

\section*{Author Contributions}
M.G. and M.J.C synthesised the samples; carried out the magnetic
measurements and the single crystal X-ray experiments; M.G., L.C.D. and
M.J.C. carried out the single crystal X-ray analysis; M.G., J.Y.L, O.F.,
L.C.D. and M.J.C. carried out the neutron diffraction experiments; M.G.,
O.F., L.C.D. and M.J.C. carried out the neutron diffraction and magnetic
property analysis; S.L. carried out the density functional calculations;
M.G., L.C.D. and M.J.C. wrote the paper with contributions from all the
authors. Research data and analysis notebooks are available at the
Nottingham Research Data Management Repository DOI:10.17639/nott.7259.



\balance


\bibliography{references.bib}
\bibliographystyle{rsc} 

\end{document}